# COMPARATIVE STUDY OF HIDDEN NODE PROBLEM AND SOLUTION USING DIFFERENT TECHNIQUES AND PROTOCOLS

Viral V. Kapadia, Sudarshan N. Patel and Rutvij H. Jhaveri

**Abstract**— Hidden nodes in a wireless network refer to nodes that are out of range of other nodes or a collection of nodes. We will discuss a few problems introduced by the RTS/CTS mechanism of collision avoidance and focus on the virtual jamming problem, which allows a malicious node to effectively jam a large fragment of a wireless network at minimum expense of power. We have also discussed WiCCP (Wireless Central Coordinated Protocol) which is a protocol booster that also provides good solution to hidden nodes.

—————————— ◆ ——————————

## 1 HIDDEN TERMINAL PROBLEM

HIDDEN nodes are the nodes that are not in the range of other nodes or a group of nodes. Take a physical star topology with an access point with many nodes surrounding it in a circular fashion: Each node is within communication range of the access point, but the nodes cannot communicate with each other as they do not have physical connection to each other. In a wireless network, it is possible that the node at the far edge of the access point's range, known as r, can see the access point, but it is unlikely that the same node can see a node on the opposite end of the access point's range, r2. These nodes are known as hidden. The problem is when nodes r and r2 start to send packets simultaneously to the access point. Since node r and r2 cannot sense the carrier, Carrier Sense Multiple Access with Collision Avoidance (CSMA/CA) does not work. To overcome this problem, handshaking is implemented in conjunction with the CSMA/CA scheme. The same problem exists in a MANET [2].

The hidden node problem can be observed easily in widespread (>50m radius) WLAN setups with many nodes that use directional antennas and have high upload. This is why IEEE 802.11 is suited for bridging the last mile, for broadband access, only to a very limited extent. Newer standards such as Wi-MAX assign time slots to individual stations, thus preventing multiple nodes from sending simultaneously and ensuring fairness, even in over-subscription scenarios [2].

IEEE 802.11 uses 802.11 RTS/CTS acknowledgment and handshake packets to partly overcome the hidden node problem. RTS/CTS is not a complete solution and may decrease throughput even further, but adaptive acknowledgments from the base station can help too.

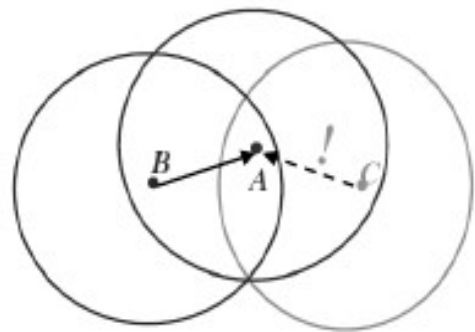

Figure. 1 Hidden Node [3]

## 2 CARRIER SENSE MULTIPLE ACCESS (CSMA)

In Carrier Sense Multiple Access:
1. If the channel is idle then transmit.
2. If the channel for communication is free then it is going to transmit without any precaution that there might be collision.
3. If the channel is busy, wait for a random time.
4. Waiting time is calculated using Truncated Binary Exponential Backoff (BEB) algorithm.

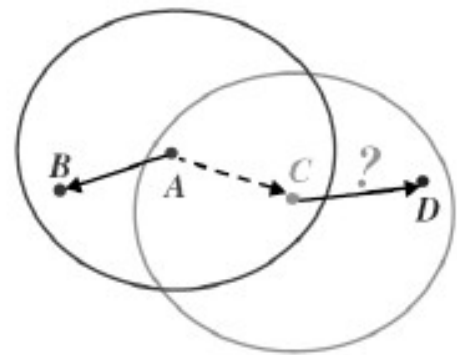

Fig. 2 Exposed Nodes [3]

————————————————
- *Viral V. Kapadia[1] is with the Department of Computer Engineering, Birla Vishvakarma Mahavidyalaya, Vallabh Vidyanagar,, Gujarat, India.*
- *Sudarshan N. Patel[2] is with the Department of Computer Engineering, A.D. Patel Institute of Technology, New Vallabh Vidyanagar, Gujarat, India.*
- *Rutvij H. Jhaveri[3] is with the Department of Computer Engineering and Information Technology, Shri S'ad Vidya Mandal Institute of Technology, Bharuch, Gujarat, India.*

## 3 HIDDEN TERMINALS

The notorious hidden node problem deals with a configuration of three nodes, like *A*, *B*, and *C* in Figure 1, whereby *B* is within the transmission range of *A* and *C*, while *C* is outside the range of *A*. In a situation like this, *C* will not be able to detect the ongoing transmission of *A* to *B* by carrier sensing and, consequently, it can inadvertently interfere with *B*'s reception of *A*'s packet [1].

The transmission range of a node A is defined as the area inside which other nodes are able to correctly receive A's packets. On the other hand, the carrier sense range of A is the area encompassing those nodes whose transmission A can perceive (carrier sense) while not necessarily being able to receive the transmitted packets [1].

Generally, it is unreasonable to assume that the two areas are always the same, e.g., the carrier sense range can be twice the transmission range [7].

Suppose that every node in Figure 1 has the same transmission range (represented by a solid circle). Node *C* is out of the transmission range of node *A* and thus would appear as a hidden node to *A*. However, if the carrier sense range of *C* is larger than the transmission range of *A* (see the dashed circle), *C* is no more hidden because it can sense the transmission of *A* and thus avoid interfering with it. This mechanism for eliminating the hidden node problem has been described in [7].

## 4 EXPOSED TERMINALS

In wireless networks, the **Exposed Node Problem** occurs when a node is prevented from sending packets to other nodes due to a neighboring transmitter. Consider an example of 4 nodes labeled R1, S1, S2, and R2, where the two receivers are out of range of each other, yet the two transmitters in the middle are in range of each other as shown in Figure 3. Here, if a transmission between node S1 and node R1 is taking place, node S2 is prevented from transmitting to node R2 as it concludes after carrier sense that it will interfere with the transmission by its neighbor node S1. However note that node R2 could still receive the transmission from node S2 without interference because it is out of range from node S1 [1].

IEEE 802.11 RTS/CTS mechanism helps to solve this problem only if the nodes are synchronized. When a node hears an RTS from a neighboring node, but not the corresponding CTS, that node can deduce that it is an exposed node and is permitted to transmit to other neighboring nodes [1]. If the nodes are not synchronized, the problem may occur that the sender will not hear the CTS or the ACK during the transmission of data of the second sender Figure 4.

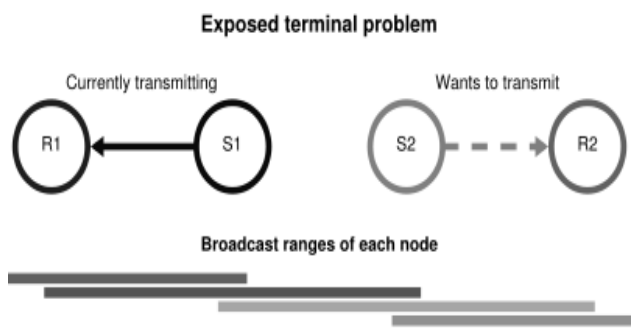

Fig. 3 Exposed Terminal Problem [3]

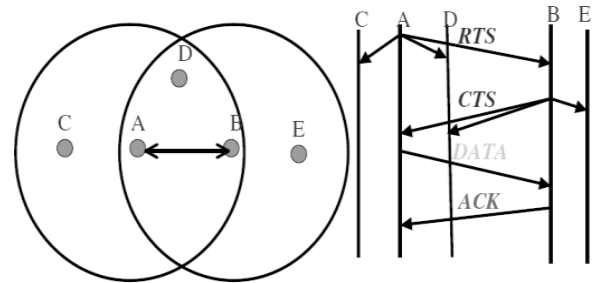

Fig. 4 RTS/CTS HANDSHAKE with ACK [6] [7]

## 5 RTS-CTS HANDSHAKE IN ACTION

- A is the source which is in the range of B, D and C.
- B is the destination which is in the range of A, D and E.
- A is the source which is in the range of B, D and C.
- B is the destination which is in the range of A, D and E.
- B sends ACK after receiving one data packet.
- Improves link reliability using ACK Figure 4.

## 6 MULTIPLE ACCESS COLLISION AVOIDANCE (MACA)

- Uses Request-To-Send (RTS) and Clear To-Send (CTS) handshake to reduce the effects of hidden terminals.
- Data transfer duration is included in RTS and CTS, which helps other nodes to be silent for this duration.
- If a RTS/CTS packet collides, nodes wait for a random time which is calculated using BEB algorithm.

Drawback:
Cannot avoid RTS/CTS control packet collisions.

## 7 SOLUTIONS

The other methods that can be employed to solve hidden node problem are:
- Increase transmitting power from the nodes.
- Use Omni-directional antennas.
- Remove obstacles.
- Move the node.
- Use protocol enhancement software.
- Use antenna diversity.
- Wireless Central Coordinated Protocol.

### 7.1 Increase Transmitting Power from the Nodes

Increasing the power (measured in mWatts) of the nodes can solve the hidden node problem by allowing the cell around each node to increase in size, encompassing all of the other nodes. This configuration enables the non-hidden nodes to detect, or hear, the hidden node. If the non-hidden nodes can hear the hidden node, the hidden node is no longer hidden. Because wireless LANs use the CSMA/CA protocol, nodes will wait for their turn before communicating with the access point.

### 7.2 Use Omni-directional Antennas

Since nodes using directional antennas are nearly invisible to nodes that are not positioned in the direction the antenna is aimed at, directional antennas should be used only for very small networks (e.g., dedicated point-to-point connections). Use Omni-directional antennas for widespread networks consisting of more than two nodes [2].

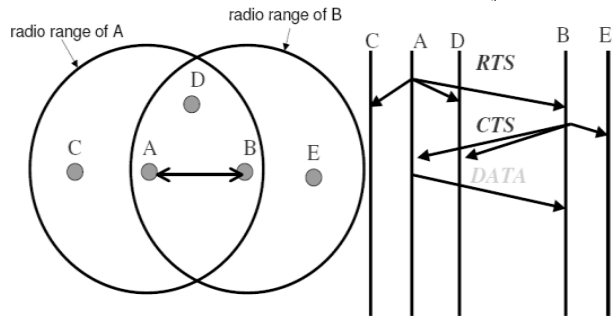

Fig. 5 RTS/CTS HANDSHAKE without ACK [6] [7]

### 7.3 Remove Obstacles

Increasing the power on mobile nodes may not work, if for example, the node that is hidden is that hiding behind a cement or steel wall preventing communication with other nodes. It is doubtful that one would be able to remove such an obstacle, but removal of the obstacle is another method of remedy for the hidden node problem. Keep these types of obstacles in mind when performing a site survey [2].

### 7.4 Move the Node

Another method of solving the hidden node problem is moving the nodes so that they can all hear each other. If it is found that the hidden node problem is the result of a user moving his computer to an area that is hidden from the other wireless nodes, it may be necessary to have that user move again. The alternative to forcing users to move is extending the wireless LAN to add proper coverage to the hidden area, perhaps using additional access points.

### 7.5 WiCCP (Wireless Central Coordinated Protocol)

WiCCP is a protocol booster for 802.11b DCF based wireless networks that provides cyclic token-passing medium access, and scheduled allocation of the available network resources, eliminating the "Hidden Node" problem. It is a pure kernel implementation resulting in high efficiency traffic control. Its not required extra configuration e.g. static ARP tables or dedicated routing contexts. WiCCP can be used in fixed wireless network deployments [9].

It is interesting to note under what conditions WiCCP will work, and when it will not work - at least optimally. WiCCP will outperform systems that do not run it when the utilization of the bandwidth increases above some high percentage. If we are running standard Ethernet utilization would be about 80%. Above this percentage of utilization, whatever that is, the channel assignment ability of WiCCP will allow the utilization to increase almost to 100% or at least as close as is humanly possible [9].

Looking on the other end of the scale, standard 802.11b will work best when the utilization is low, and the levels are set correctly so that at the access point all power level are the same. Under low utilization it is likely that the power levels do not affect things too much.

The main question is regarding heavy traffic. WiCCP allows a guarantee of bandwidth for a particular user and this solution appears to be the correct solution for this case to solve this problem. The ability to offer a guarantee and then offer more on top of that where available is worthwhile. The problem is the overhead of polling.

As the number of users increases, WiCCP will tend to have issues with assigning timeslots to each, ensuring latency. Standard 802.11b will have a definite advantage when there are a lot of stations, and very few want to transmit most of the time - Such as 50 laptops who only check their mail once every 15 minutes (without reading), as opposed to 50 users attempting to surf the web.

## 8. CONCLUSION

Hidden node problem can be solved by many means but each solution is for particular scenario. Using different techniques like Increase Transmitting Power From the Nodes, Use Omni-directional antennas, Remove obstacles, Move the nodes, Use protocol enhancement software, Use antenna diversity, Wireless Central Coordinated Protocol etcetera would increase the performance of ad-hoc networks a lot.

**Viral V. Kapadia** – Ph.D. candidate, Lecturer, Department of Computer Engineering, Birla Vishvakarma Mahavidyalaya, Vallabh Vidyanagar. He is author of 6 papers, with 3 papers in international conferences and 3 in national conferences.

**Sudarshan N. Patel** – Ph.D. candidate, Lecturer, Department of Computer Engineering, A.D. Patel Institute of Technology, New Vallabh Vidyanagar. The research area of interest are Mobile Ad-hoc Networks and Distributed Operating System.

**Rutvij H. Jhaveri** – Member of ISTE, Sr. Lecturer, Department of Computer Engineering, Shri S'ad Vidya Mandal Institute of Technology, He is author of 5 papers, with 1 paper in international conference and 4 in national conferences. The research area of interest is: "Mobile Ad-hoc Networks".